\documentclass[a4paper,conference]{IEEEtran}

\IEEEsettopmargin{t}{30mm}
\IEEEquantizetextheight{c}
\IEEEsettextwidth{14mm}{14mm}
\IEEEsetsidemargin{c}{0mm}

\usepackage[utf8]{inputenc}
\usepackage[T1]{fontenc}
\usepackage{url}
\usepackage{ifthen}
\usepackage{cite}
\usepackage[cmex10]{amsmath}

\usepackage{amsthm}
\DeclareMathOperator*{\argmax}{arg\,max}

\usepackage{algorithmic, algorithm}
\usepackage[english]{babel}
\usepackage[table]{xcolor}
\usepackage{enumitem}
\usepackage{amssymb}
\usepackage{xcolor}
\usepackage{graphicx}

\usepackage{subcaption}  
\usepackage{tikz}
\usetikzlibrary{positioning, fit, shapes.geometric, arrows}

\tikzstyle{block} = [rectangle, draw, minimum height=0.5cm, text centered, text width=4cm]
\tikzstyle{arrow} = [thick,->,>=stealth]
\tikzstyle{dashedblock} = [draw, dashed, rounded corners, inner sep=0.5em, minimum width=4cm]

\usepackage[font=normalsize]{caption}
\usepackage{comment}

\theoremstyle{definition}

\title{\huge DeepDIVE: Optimizing Input-Constrained Distributions \\for Composite DNA Storage via Multinomial Channel}

\author{\IEEEauthorblockN{Adir Kobovich, Eitan Yaakobi and Nir Weinberger}
\IEEEauthorblockA{\textit{Technion -- Israel Institute of Technology, Haifa, Israel} \\
Email: adir.k@campus.technion.ac.il, yaakobi@cs.technion.ac.il, nirwein@technion.ac.il}
}

\IEEEoverridecommandlockouts
\IEEEaftertitletext{\vspace{-5pt}}

\begin{document}
\maketitle

\begin{abstract}
We address the challenge of optimizing the capacity-achieving input distribution for a multinomial channel under the constraint of limited input support size, which is a crucial aspect in the design of DNA storage systems. We propose an algorithm that further elaborates the Multidimensional Dynamic Assignment Blahut-Arimoto (M-DAB) algorithm \cite{Kobovich2024MDAB}. Our proposed algorithm integrates variational autoencoder for determining the optimal locations of input distribution, into the alternating optimization of the input distribution locations and weights. 
\end{abstract}


\section{Introduction}
\label{sec:introduction}
DNA storage is a rapidly advancing technology that encodes digital data into sequences of nucleotides using quaternary encoding, where the bases $A$, $C$, $G$, and $T$ represent the information \cite{church2012next, goldman2013towards}. These sequences, or \textit{strands}, are produced through a process called \textit{synthesis} and retrieved via \textit{sequencing}. A key aspect of this method is the generation of multiple copies of each strand during synthesis. In this paper, we explore a novel approach to utilizing this redundancy by introducing \textit{composite DNA letters} \cite{anavy2019data, choi2019high, preuss2021data, zhang2022limited, yan2023scaling, Kobovich2024MDAB}. Composite DNA letters are formed by mixing different nucleotides and have been shown to improve data encoding performance in experiments \cite{anavy2019data, choi2019high, yan2023scaling}. The potential benefits are significant: while standard four-letter DNA encoding is limited to $\log(4) = 2$ bits per channel use, composite encoding offers an unbounded capacity, enabling shorter strands to encode more data. This is crucial because shorter strands reduce synthesis costs \cite{choi2019high} and lower the risk of errors, which increase with strand length \cite{bornholt2017toward}.
Writing a composite letter and reading $n$ copies randomly can be modeled as a noisy communication channel, in particular as a \textit{multinomial channel} \cite{Kobovich2024MDAB}. The input to this channel is a probability vector of length $k=4$, representing a mixture of nucleotides. The channel output follows a multinomial distribution, with $n$ trials and probabilities determined by the input vector.
The channel's maximum information storage rate, or capacity, is obtained by maximizing the mutual information between the input and output, over all feasible choices of input distributions \cite{cover1999elements}, that is, distributions over the $(k-1)$-dimensional probability simplex. 
Previous work \cite{Kobovich2024MDAB} has shown that even for small values of $n$ (e.g., $n=9$), the input distribution that maximizes capacity requires dozens of mass points. Furthermore, as indicated by the scaling law \cite{abbott2019scaling}, the support size grows exponentially with the capacity. This presents a challenge for DNA storage systems, where each mass point corresponds to a distinct nucleotide mixture, and the number of possible mixtures is limited. 
To address this issue, our paper focuses on calculating a capacity-achieving input distribution for the multinomial channel, subject to a constraint on the support size.
We follow a well-established decomposition of the problem of finding the optimal input distribution, to alternating between determining the weights of the mass points and their locations. Our approach is based on deep learning and introduces a novel way to discretize the multinomial channel, providing valuable insights into the characteristics of the capacity-achieving input distribution and achieving significant improvements in the DNA storage domain.
The paper is organized as follows. Section~\ref{sec:Problem Definition} introduces the input-constrained multinomial channel optimization problem. Section~\ref{sec:RelatedWork} reviews prior works on the multinomial channel and autoencoder-based communication systems. Section~\ref{sec:method} describes our proposal for a neural network architecture and an optimization procedure. Section~\ref{sec:results} presents the input distribution and the corresponding channel capacities achieved by our method. Finally, Section~\ref{sec:conclusion} concludes the paper.

\section{Problem Statement}
\label{sec:Problem Definition}
This section provides a formal definition of the input-constrained multinomial channel, along with the associated optimization problem for the capacity-achieving input distribution (CAID). The input alphabet of the multinomial channel is the $(k-1)$-dimensional probability simplex, denoted as $\Delta_{k}:=\{x\in\mathbb{R}_{+}^{k}\mid\sum_{i=1}^{k}x_{i}=1\}$.
When $n$ samples are available, the output alphabet comprises all multisets of cardinality $n$ derived from the set $[k]$, represented as $\mathcal{Y}_{n,k}:=\{y\in\mathbb{Z}_{+}^{k}\mid\sum_{i=1}^{k}y_{i}=n\}$. Its cardinality given by $|\mathcal{Y}_{n,k}| = \binom{n+k-1}{k-1}$. 
For an input $x \in \Delta_k$, the output $Y$ of the multinomial channel follows a multinomial distribution, denoted as $Y\sim \mathrm{Multinomial}(n,x)$.
That is, the transition probability for obtaining the output $y$ given the input $x$ is 
\begin{equation} \label{eq: multinomial channel}
 P_{Y|X}^{(n,k)}(y|x)= \frac{n!}{\prod_{j = 1}^{k} y_j!} \prod_{j = 1}^{k}x_j^{y_j}.
\end{equation}
Thus, for an input letter $x\in\Delta_k$, the expected occurrence of the $i$-th letter in the output strand is given by $nx_i$.

This channel model captures only the randomness of the output resulting from the sampling of the input, and does not account for any additional noise during the reading process. If we further assume that the reading process can be modeled as a symmetric discrete memoryless channel (DMC) with a total flip probability of $\epsilon$ (thus distributing $\frac{\epsilon}{k-1}$ 
to each of the other $k-1$ letters), the model is modified to a $\mathrm{Multinomial}(n,x \ast \epsilon)$ channel, where 
\begin{equation}
   (x \ast \epsilon)_i := x_i(1-\epsilon)+\epsilon(1-x_i) \hspace{3mm} \text{for all} \hspace{3mm} i\in [k]. 
\end{equation}
Therefore, our algorithm and findings can be easily extended to accommodate this scenario. For simplicity, we will primarily focus on the noiseless channel in the subsequent discussions.

It has been established in \cite{Kobovich2024MDAB} that a CAID exists with a finite support size $m \leq |\mathcal{Y}(k,n)|$, and so the corresponding input distribution can be expressed as using the Dirac delta function $\delta(x)$ as 
\begin{equation} \label{eq: atomic input distribution} 
\mathnormal{f}^*_X(x)=\sum\limits_{i = 1}^{m} p^*_{i} \delta(x-x^{(i)}).
\end{equation}
Consequently, ${f}^*_X(x)$ is an atomic distribution. 
We refer to the adjustment of the weights as \textit{probabilistic shaping}, and that of the locations \textit{geometric shaping} \cite{stark2019joint}. 

Our primary objective is to determine the capacity of the channel under the constraint that the input distribution is supported on at most $d$ atoms, where $d<m$. Thus, our goal is to solve the following optimization problem to identify a CAID of the input-constrained multinomial channel, expressed as
\begin{equation}
C_{n,k,d} := \max\limits_{f_X\in\mathcal{F}_{k, d}} I(X;Y),
\label{eq: capacity optimization problem}
\end{equation}
where $\mathcal{F}_{k, d}$ be the set of all atomic input distributions supported on the input alphabet $\Delta_k$ with support size $d$.

\section{Related Work}
\label{sec:RelatedWork}
\subsection{The Multinomial Channel and Capacity Optimization}
The simpler case of input dimension $k=2$ is known as the \textit{binomial channel} \cite{komninakis2001capacity}. In \cite{wesel2018efficient}, an algorithm for its input optimization, called, the \textit{Dynamic Assignment Blahut-Arimoto} (DAB) algorithm,  was introduced.  
DAB operates as a primal-dual alternating optimization algorithm, alternating between finding optimal weights for fixed locations and optimizing locations for given weights.
When the locations are fixed, the channel simplifies to a discrete memoryless channel, allowing the classical Blahut-Arimoto algorithm \cite{blahut1972computation, arimoto1972algorithm} to compute the optimal input probabilities.
To identify the optimal locations, DAB leverages the capacity dual optimization problem' also known as the Csiszár minimax capacity theorem \cite{csiszar2011information}.
Later, in \cite{Kobovich2024MDAB}, the DNA storage channel using composite symbols was modeled as a multinomial channel, adapting DAB to the multidimensional case (M-DAB). A key adjustment was limiting the search space to functions that exhibit symmetry under any permutation of dimensions. 
However, in this paper we consider a multinomial channel with an additional constraint on the input support size. For this case, the Csiszár minimax capacity theorem no longer provides a tight capacity bound, and the capacity-achieving input distribution is not guaranteed to retain such symmetry. Consequently, the problem does not seem to be tractable to solve, while solely relying on expert-based approaches. 

\subsection{Deep Learning in Channel Coding}
In general, traditional methods often fall short when addressing complex optimization problems, particularly in large or non-convex search spaces. Directly solving these problems becomes intractable due to the vast number of potential input configurations and the intricate nature of the objective function. As discussed above, our problem falls into this category. Nonetheless, it turned out that recent advances in deep learning can be applied to
solving such problems, and specifically, it has been applied to constellation design in communication systems; see \cite{lopez2020survey} for a survey. 

A prominent method in this domain is \textit{end-to-end learning}, introduced in \cite{o2017introduction}. This approach focuses on optimizing transmitter and receiver designs for specific performance metrics and channel models, treating the entire communication chain as an autoencoder, a form of unsupervised learning. The use of this method for jointly learning both geometric and probabilistic constellation shaping is demonstrated in \cite{stark2019joint}. The primary limitation of this method stems from the requirement to train the entire autoencoder, as the channel must be modeled as a neural network, necessitating its differentiability. 
To facilitate the training of communication systems with unknown channel models or non-differentiable components, previous studies have sought to learn approximations of the channel using Generative Adversarial Networks (GANs) \cite{ye2018channel}, or Reinforcement Learning \cite{aoudia2018end}. Another approach utilizes a neural network to estimate mutual information \cite{belghazi2018mine}, followed by maximizing the output of this estimator \cite{fritschek2019deep}.

\begin{figure}[h!]
\begin{center}
\vspace{-10mm}
\begin{tikzpicture}[node distance=1.7cm]

\node (input) [] {};

\node (dense1) [block, below of=input] {Dense Neural Network};
\node (norm) [block, below of=dense1, node distance=0.5cm] {Softmax};
\node (encoder) [dashedblock, fit={(dense1) (norm)}, label={[anchor=north west, yshift=5mm]north west:Encoder}, label={right:$ f_{\theta_E} (\cdot)$}] {};

\node (sampler) [block, below of=norm] {Gumbel Sampler};
\node (softmax_t) [block, below of=sampler, node distance=0.5cm] {Softmax $\tau$};
\node (norm1) [block, below of=softmax_t, node distance=0.5cm] {Normalization};
\node (channel) [dashedblock, fit={(sampler) (softmax_t) (norm1)}, label={[anchor=north west, yshift=5mm]north west:Channel}, label={right:$p(y|x)$}] {};

\node (dense2) [block, below of=channel,node distance=2.2cm] {Dense Neural Network }; 
\node (softmax) [block, below of=dense2, node distance=0.5cm] {Softmax};
\node (decoder) [dashedblock, fit={(dense2) (softmax)}, label={[anchor=north west, yshift=5mm]north west:Decoder}, label={right:$ f_{\theta_D} (\cdot)$}] {} ;

\node (output) [below of=softmax, minimum height=0cm] {};

\draw [arrow] (input.south) ++(0, -5mm) -- (dense1.north) node[midway, right, yshift=1mm] {$s \in \mathcal{S}$};
\draw [arrow] (norm.south) -- (sampler.north) node[midway, right] {$x \in \Delta_k$};
\draw [arrow] (norm1.south) -- (dense2.north) node[midway, right] {$\tilde{y} \in \Delta_k$};
\draw [arrow, shorten >=5mm] (softmax.south) -- (output.north)  node[midway, right, yshift=2mm] {$\hat{s} \in \Delta_d$};

\end{tikzpicture}
\end{center}
\vspace{-10mm}
    \caption{End-to-end autoencoder model.}
    \label{fig:architecture}
\vspace{-10pt}
\end{figure}
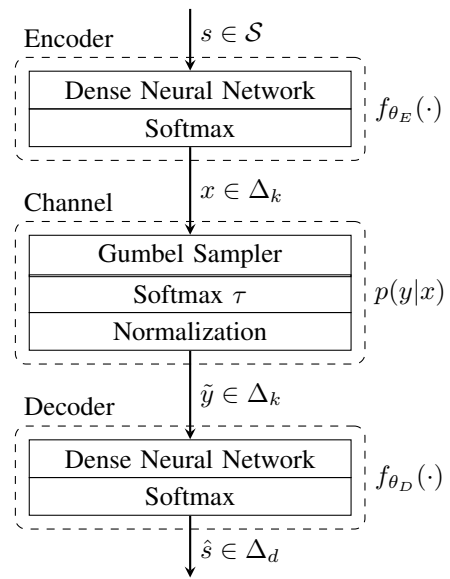

\section{Proposed Method}
\label{sec:method}
As discussed, given the complexity of the optimization problem, we turn to deep learning techniques.   In particular, we introduce an alternating optimization algorithm that combines the Blahut-Arimoto algorithm for determining the weights of the $d$ atoms (for given $d$ locations), and a Variational Autoencoder (VAE)~\cite{kingma2013auto}, to identify the optimal $d$ locations of the mass points (for given $d$ weights). In this way, the effectiveness of the expert-based knowledge is expressed through the use of an alternate minimization algorithm and the Blahut-Arimoto algorithm. The use of neural network is limited to the  parts which cannot be addressed by principled methods, to wit, the optimization of the locations (geometric shaping). 

The multinomial channel we consider is non-differentiable. Such situations are often addressed by model-free methods, yet we opt for an alternative approach, which avoids the need for the model to learn the channel itself, which can be both challenging and inefficient. To handle the discrete and non-differentiable nature of the channel output, we employ the Gumbel-Softmax trick~\cite{jang2016categorical}, which provides a differentiable approximation for sampling. In the next sections, we further detail the architecture of the Variational Autoencoder model (see Figure~\ref{fig:architecture}) and outline its learning procedure.

\subsection{Variational Autoencoder Architecture}
Each channel symbol $s$ is represented as a one-hot vector of size $d$, such that $s \in \mathcal{S} = \{ e_i \mid i = 1, \ldots, d\}$ where $e_i$ has a value of $1$ at position $i$ and $0$ elsewhere. 
The encoder, denoted as $ f_{\theta_E} (\cdot)$, consists of a single hidden layer with $256$ units and uses $\mathtt{ReLU}$ activation function~\cite{nair2010rectified}. The output layer has a dimensionality of  $k$ and applies the softmax function~\cite{goodfellow2016deep}, ensuring that the channel input $x$ lies within the simplex $\Delta_k$. 

While previous works primarily focus on the AWGN channel, where the reparametrization trick is directly applicable, we employ the Gumbel-Softmax trick to facilitate sampling from the Multinomial channel. 
We enumerate the elements of the channel output set $\mathcal{Y}_{n,k}$ and denote by $p(i|x)$ the probability of observing the $i$th element in this enumeration, given the input $x$. Specifically, the probability is given by:
\begin{equation}
    p(j|x) = \Pr \left( j = \argmax\limits_{i=1, \ldots, |\mathcal{Y}_{n,k}|} \left(g_i + \log p(i|x)\right) \right), 
\end{equation}
where $g_1, \ldots, g_k$ are independent samples drawn from the standard Gumbel distribution, characterized by the probability density function:
\begin{equation}
f(x) = e^{-(x + e^{-x})}.
\end{equation}
To approximate the $\argmax$ operation in a differentiable manner, we utilize the softmax function. This produces a distribution vector, which provides a smooth approximation of the one-hot vector representation of the output sample:
\begin{equation}
    p(j|x) \approx 
    \frac{  \exp{( g_j + \log p(j|x)) / \tau } }
        {\sum\limits_{i=1}^{|\mathcal{Y}_{n,k}|} 
        \exp{ ( g_i + \log p(i|x) ) / \tau }
    } , \hspace{3mm} j=1, \ldots, |\mathcal{Y}_{n,k}|,
\end{equation}
where $\tau > 0$ is a temperature parameter controlling the degree of approximation to the $\argmax$. In our model we fix $\tau = 0.01$. 
The final step in the channel process is to convert the distribution vector back to a numerical representation. This is achieved by calculating the expected trials outcomes based on the distribution and then normalizing the result as follows:
\begin{equation}
    \tilde{y} = \frac{1}{n} \mathop{\mathbb{E}}_{p(\cdot |x) } \left[ j \right]
\end{equation}
yielding a probability vector $\tilde{y} \in \Delta_k$, which represents the proportion of each trial result.
The decoder, denoted as $ f_{\theta_D} (\cdot)$, follows a similar structure to the encoder. It consists of a single hidden layer with $256$ units and uses the $\mathtt{ReLU}$ activation function.
The output layer has a dimensionality of $d$ and applies the softmax function, producing $p_{\theta_D}(s|y)$. The decoded message $\hat{s}$ is then determined as the index of the element in $p$ with the highest probability.

\subsection{Training Procedure}
The model is trained end-to-end using stochastic gradient descent (SGD)~\cite{goodfellow2016deep}, specifically utilizing the Adam optimizer~\cite{KingmaB14}, on the set of possible messages.
Due to the stochastic nature of the channel, a large batch size is necessary.  While the number of symbols is relatively small, we repeat the symbols to define the epoch size. 
In our implementation, we use a batch size of $32,768$ and an epoch size of $1,048,576$, with the number of epochs ranging from $150$ to $300$.

The training objective is to minimize the categorical cross-entropy loss~\cite{goodfellow2016deep}
\begin{equation}
\mathcal{L}(\theta_E, \theta_D)  \triangleq \mathbb{E}_{s,y} \{ - \log \left(  \tilde{p}_{\theta_D}(s|y) \right) \}.
\end{equation}

\begin{figure}[t]
  \centering
\includegraphics[width=1\linewidth,keepaspectratio]{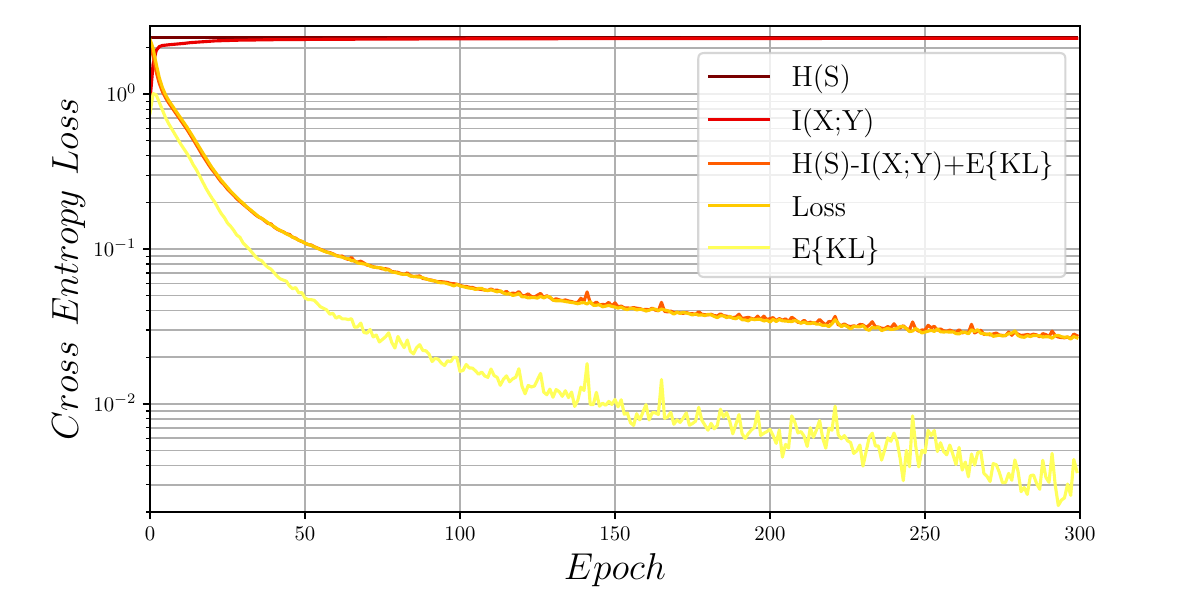}
  \caption{Loss decomposition during the training process. } 
  \label{fig:LossDecomposition}
\vspace{-10pt}
\end{figure}

By denoting the channel input distribution as $p_S$, we can express the following decomposition~\cite{stark2019joint}
\begin{align}
    \mathcal{L}(\theta_E, \theta_D) =& H_{p_S}(S) - I_{p_S, \theta_E}(X;Y) \notag\\ +& \mathbb{E}_y\{ D_{KL} ( p_{p_S,\theta_E}(x|y)||p_{\theta_D}(x|y)) \},
\end{align}
where $D_{KL}(\cdot||\cdot)$ denotes the Kullback–Leibler (KL) divergence.
From this, we conclude that minimizing the cross-entropy effectively maximizes the mutual information, which is our primary objective, while introducing a penalty term associated with the decoder's approximation of the true posterior distribution $p_{p_S,\theta_E}(s|y)$. 
Figure~\ref{fig:LossDecomposition} illustrates an example of the loss during the training process and highlights its decomposition. Notably, the penalty term is negligible. 
During the training process, we opted to use a weighted cross-entropy loss, where the weights are determined by $p_S$, rather than relying on sampling. At the end of each epoch, the weights were updated using the Blahut-Arimoto algorithm. 

\section{Experimental Results}
\label{sec:results}
In this section, we present the results of our model, which introduces a novel approach to discretizing the multinomial channel—a method that, to the best of our knowledge, has not been explored before. Due to the lack of direct benchmarks for comparison, we begin by evaluating our model against other techniques applicable only in the one-dimensional case of the binomial channel.
We then highlight a key insight from our results: using the simplex vertices is not always optimal. Finally, we apply our model to the DNA storage domain, comparing its composite symbols to those used in previous experiments. Our approach demonstrates a significant improvement in performance.

\subsection{Binomial Discretization}

\begin{figure}[t]
  \centering
\includegraphics[width=1\linewidth,keepaspectratio]{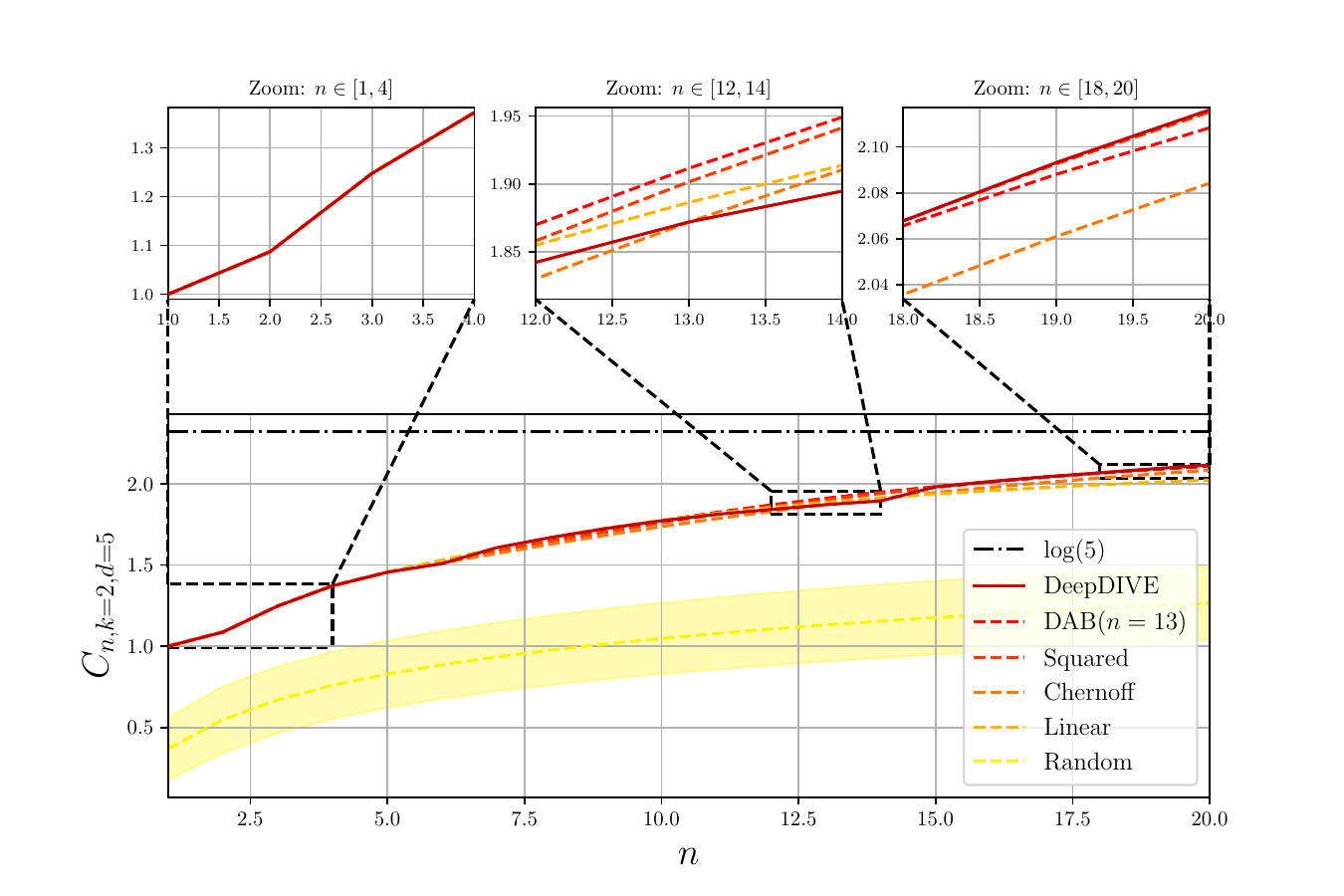}
  \caption{DeepDIVE's geometric and probabilistic shaping results compared to previous methods.}
  \label{fig:BA20}
\vspace{-10pt}
\end{figure}

We evaluated the result of our model result by comparing it with different discretization methods of the one-dimensional probability simplex $\Delta_{2}:=\{ (x, 1-x) \mid x \in [0, 1]\}$. The results are shown in Figure~\ref{fig:BA20}.
The only method easily generalized to multidimensional simplex is to use random symbols (labeled as \emph{Random}), specifically drawn from $\mathrm{Dirichlet}(1, \ldots, 1)$.
Although straightforward, this method often yields suboptimal results because it does not account for the distances between symbols.
A simple alternative is to separate the symbols by equal distances. This results in a linear support of the form:
\begin{equation}
x \in \left\{0, \frac{1}{d-1}, \ldots, \frac{d-2}{d-1}, 1\right\},
\end{equation}
which we label as \emph{Linear}. While this method ensures uniform spacing, it does not account for the varying influence that each symbol may have on the output.
To address these limitations, we adopt a \textit{companding approach} inspired by the fact that, without an input constraint, the CAID of the channel is asymptotically proportional to Jeffrey's prior~\cite{clarke1994jeffreys}
\begin{equation}
f(x) = \frac{1}{\pi \sqrt{x (1-x)}}.
\end{equation}
This suggests applying the following transformation
\begin{equation}
    x' = \operatorname{sign}(x - 0.5) \cdot \sqrt{\frac{\lvert x - 0.5 \rvert}{2}} + 0.5,
\end{equation}
which redistributes the symbols non-linearly. The results of this transformation are labeled as \emph{Squared}.

Another discretization approach, proposed by~\cite{chang1990numerical}, suggests using the KL-divergence as a distance measure\footnote{Since the KL-divergence is not symmetric, the method identifies two sets of points which can be interpreted as centroid and boundaries}. This approach has been further studied in the context of divergence covering~\cite{tang2022divergence}.  Notably, for exponential families~\cite{nielsen2013information}, using the KL-divergence aligns with the Chernoff distance, which is widely employed in hypothesis testing~\cite{cover1999elements}. 
This method, labeled as \emph{Chernoff}, is asymptotically optimal but tends to underperform for smaller values of $n$. 
The final method we evaluate uses the CAID of the unconstrained multinomial channel for $n=13$, which is supported on $5$ symbols. This CAID is computed using DAB algorithm~\cite{wesel2018efficient} and is labeled as \emph{DAB$(n=13)$}.
Additionally, we include the theoretical upper bound of $\log(5)$, representing the maximum achievable mutual information with five symbols. 
As shown in Figure~\ref{fig:BA20}, all methods achieve capacity for small $n$, with DAB being optimal for $n = 13$, as expected. However, for larger values of $n$, our proposed method outperforms the others, demonstrating its superiority in these regimes.  

\begin{figure}[t]
  \centering
\includegraphics[width=1\linewidth,keepaspectratio]{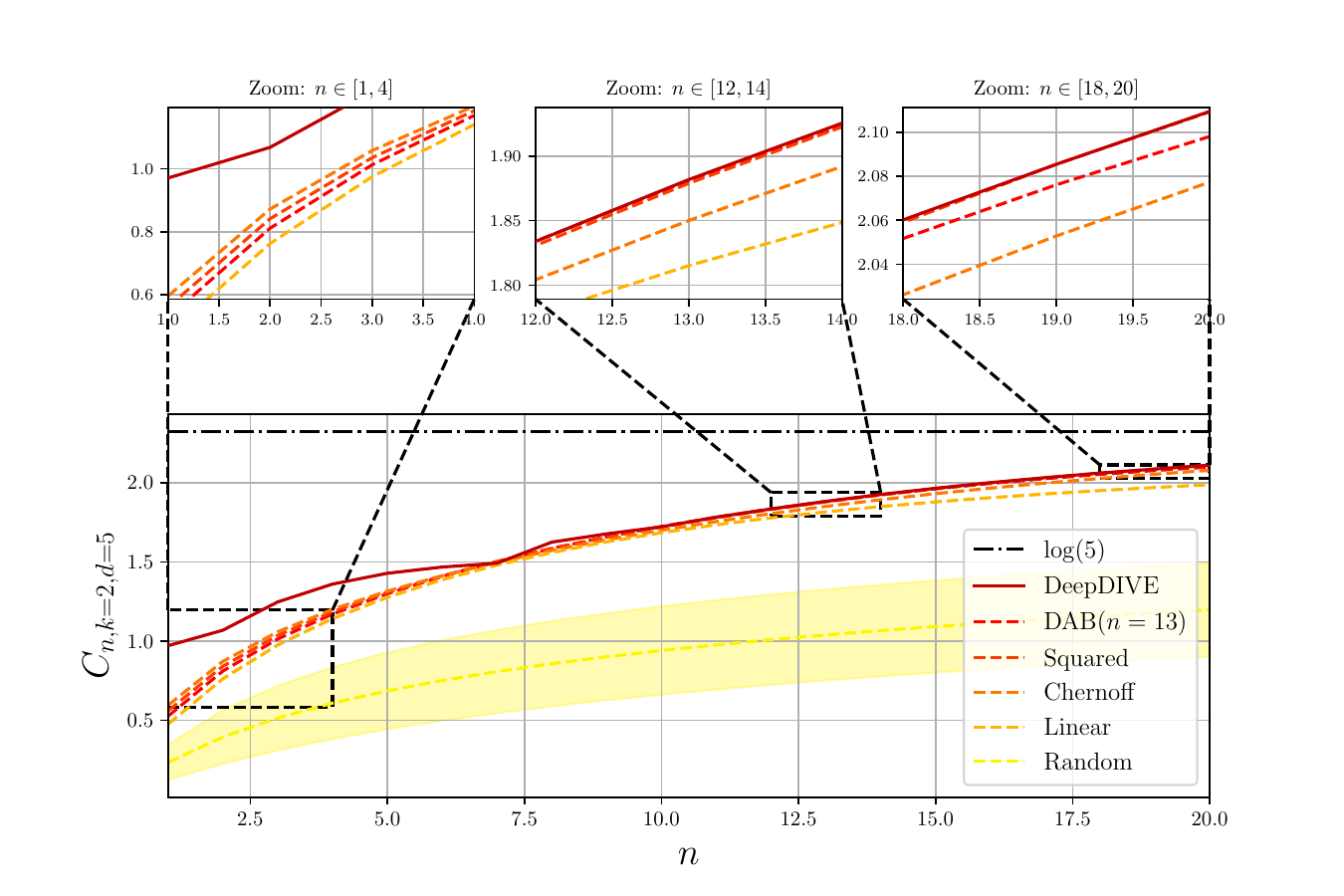}
  \caption{DeepDIVE's only geometric shaping results compared to previous methods.}
  \label{fig:fixed20}
\vspace{-10pt}
\end{figure}
Our main interest is in calculating the CAID of the input-constrained multinomial channel; therefore, we are using both geometric and probabilistic shaping, but many applications may consider equal input probabilities. This discussion can be interpreted as an average-case versus worst-case metric. 
To illustrate such an application, consider our main use-case motivation of DNA storage, where we utilize the multiple copies of each strand during the synthesis process. The capacity of the channel is achieved when the number of channel usages approaches infinity. Due to the nature of the process and to allow random access, one may prefer a coding mechanism applicable in the regime corresponding to channel usages which are equal to the strand length.

This approach, which involves only geometric shaping, is easily implemented in our framework by using cross-entropy with fixed weights. The results are presented in Figure~\ref{fig:fixed20}, and show that our model surpasses all the other discretization methods. Note that the other discretization methods do not consider the probability shaping, implying that their better results on the average case are not robust and are not likely to be generalized to the multidimensional case. Another unique feature of our approach is for the regime of small $n$ values, where the CAID support is less than the constraint; when using probability shaping the probabilities of the redundant symbols are equal to zero, disregarding them. In contrast, our method results with multiple copies of symbols allowing it to achieve a large margin over different methods.   

\subsection{Pure Symbols Non-Optimality}

It has been established that, in the binomial case, the simplex vertices (i.e. $\{0,1\}$) are part of the support for the atomic CAID\cite[Proposition 5]{barletta2024binomial}. 
While no analogous proof exists for the multinomial case, it might seem intuitive to favor the simplex vertices since these non-composite symbols introduce no randomness and always produce the same channel output.
However, for input-constrained multinomial channels, our model reveals that such a configuration is not always capacity-achieving. In hindsight, the intuition behind this result lies in the potential benefit of spacing the symbols more evenly. Moving toward the simplex edges may bring the symbols closer together, which can reduce capacity. 
  
\begin{figure}[htbp]
    \centering
    \begin{subfigure}[b]{0.49\columnwidth}
        \includegraphics[width=\linewidth,keepaspectratio]{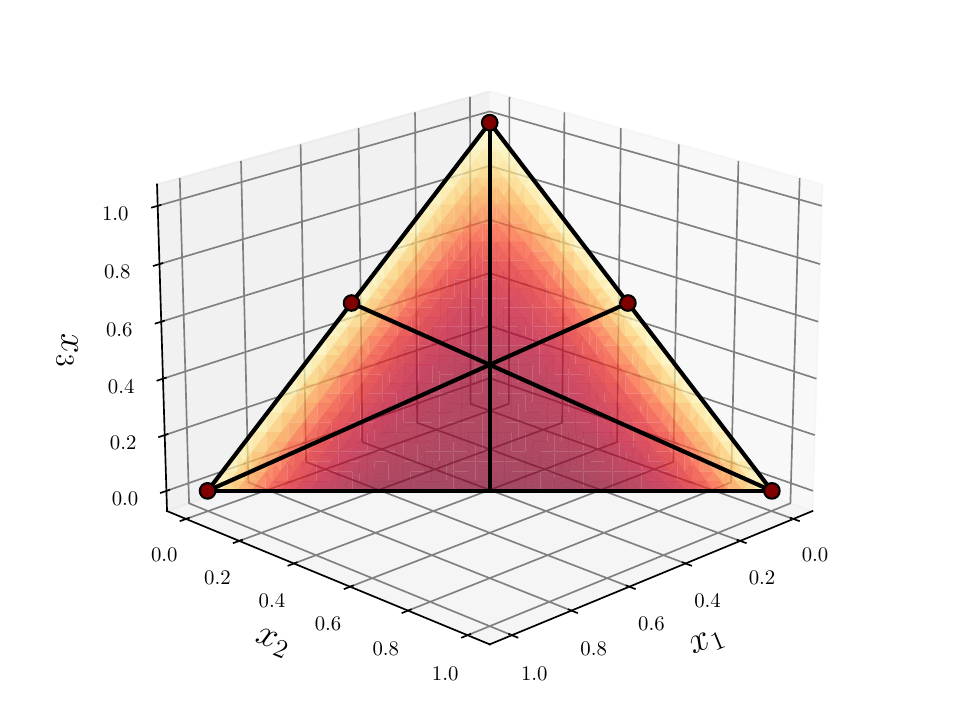}
        \caption{} 
        \label{fig:corners}
    \end{subfigure}
    \hfill
    \begin{subfigure}[b]{0.49\columnwidth}
        \includegraphics[width=\linewidth,keepaspectratio]{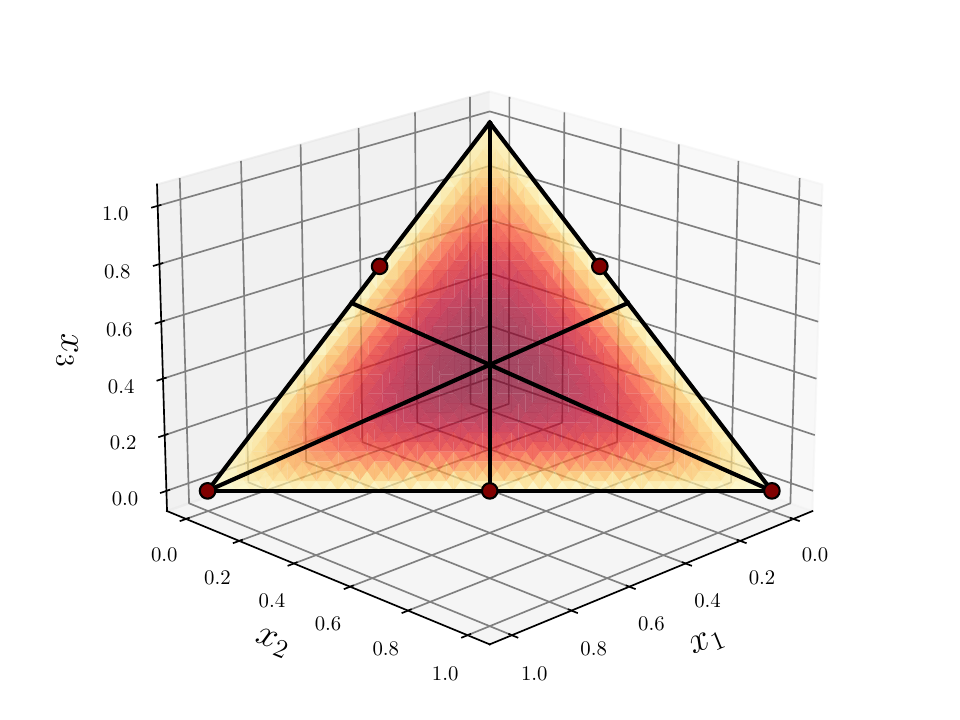}
        \caption{}  
        \label{fig:middle}
    \end{subfigure}
    \caption{Five-symbols constellations on two-dimensional simplex; corners configuration (a) and middle configuration (b).  }
    \label{fig:combined}
\end{figure}

\begin{figure}[t]
  \centering
\includegraphics[width=1\linewidth,keepaspectratio]{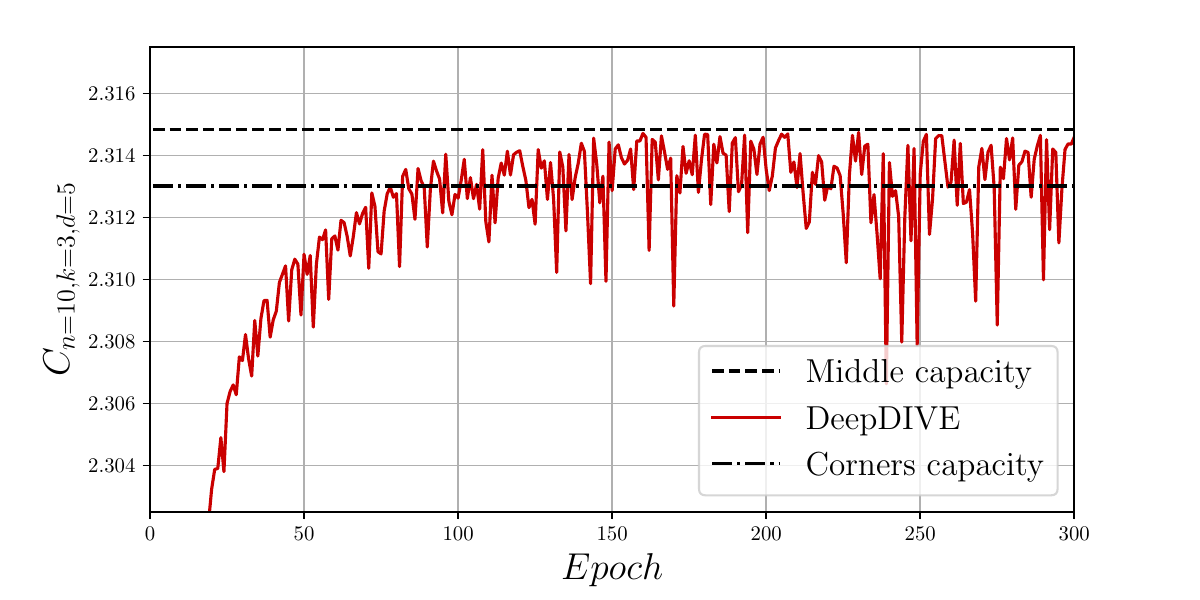}
\vspace{-15pt}
  \caption{DeepDIVE's configuration result during the training process, compared to corners and middle configuration. }
  \label{fig:5support}
\vspace{-10pt}
\end{figure}

Interestingly, an example of such a configuration arises during the training of our model for $C_{n=10,k=3,d=5}$. The $50$th iteration shows a learned constellation containing all the simplex vertices, referred to as the \emph{corners configuration} (Figure~\ref{fig:corners}). By the $100$th iteration, the constellation evolves to exclude one vertex, forming the \emph{middle configuration} (Figure~\ref{fig:middle}).
Further examination suggests that the capacity of the middle configuration is strictly larger than that of the corners configuration. The mutual information achieved by the model throughout the training is shown in Figure~\ref{fig:5support}.

\subsection{Multinomial Results}

\begin{figure}[t]
  \centering
\includegraphics[width=1\linewidth,keepaspectratio]{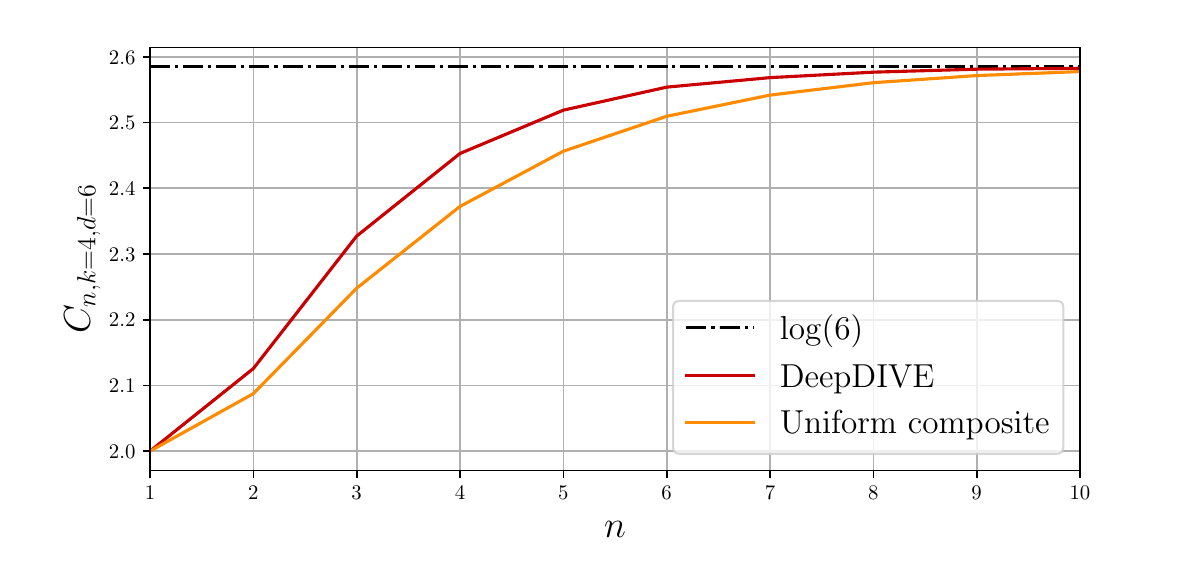}
\vspace{-15pt}
  \caption{Six-letter composite alphabet comparison.}
  \label{fig:4d6}
\vspace{-10pt}
\end{figure}

We finally present results relevant to our DNA storage application, specifically for the case $k=4$. Recent approaches, such as those in~\cite{preuss2024efficient}, propose using shortmers as the fundamental components of composite symbols. The need for $k$ larger than $4$ further underscores the significance and utility of our algorithm, which can efficiently define the CAID for arbitrary dimensions.
 In the experiment conducted by~\cite{anavy2019data}, 
a six-letter composite alphabet was used. We refer to the naive method employed as \textit{uniform composite}, which is composed of the simplex's four vertices (the pure symbols) and two composite symbols: $(0.5,0.5,0,0)$ and  $(0,0,0.5,0.5)$. However, using our method, we found that better performance can be achieved with alternative composite symbols, such as $(0.4,0.2,0.4,0)$ and $(0,0.4,0.2,0.4)$. Figure~\ref{fig:4d6} compares the mutual information achieved using our model with that of the uniform composite, showing a significant improvement.

\section{Conclusion}
\label{sec:conclusion}
In this paper, we introduced a Variational Auto-Encoder (VAE)-based alternating optimization approach to solve the constrained-input multinomial channel problem. Our method offers a novel and effective way to optimize input distributions under a support size constraint, with significant implications for DNA storage systems. By combining deep learning with the Blahut-Arimoto algorithm, we address challenges in high-dimensional input spaces while maintaining the constraint on support size.

\section*{Acknowledgments}
The research was Funded by the European Union (ERC, DNAStorage, 101045114 and EIC, DiDAX 101115134). Views and opinions expressed are however those of the authors only and do not necessarily reflect those of the European Union or the European Research Council Executive Agency. Neither the European Union nor the granting authority can be held responsible for them. The research of N. W. was supported by the Israel Science Foundation (ISF), grant no. 1782/22.

\IEEEtriggeratref{19}
\bibliographystyle{IEEEtran}

\end{document}